\documentclass[preprint,notitlepage,floatfix,aps,prd,showpacs,nofootinbib,amsmath,amssymb,amsfonts,superscriptaddress]{revtex4-1}

\usepackage{bm}
\usepackage{graphicx}
\usepackage{physics}
\usepackage{dsfont}
\usepackage[normalem]{ulem}
\usepackage{hyperref,url}
\usepackage{mathrsfs}
\usepackage{calrsfs}
\usepackage{tikz}

\usepackage[caption=false]{subfig}

\usepackage{varioref}
\usepackage[active]{srcltx}

\expandafter\ifx\csname package@font\endcsname\relax\else
\expandafter\expandafter
\expandafter\usepackage
\expandafter\expandafter
\expandafter{\csname package@font\endcsname}%
\fi
\hypersetup{
	colorlinks=true,       
	linkcolor=blue,          
	citecolor=blue,        
}

\usepackage[section]{placeins}

\usepackage{fancybox}
\labelformat{figure}{Fig.~#1}

\begin{document}
\begin{titlepage}
\pagestyle{empty}
\baselineskip=21pt
\vspace{2cm}
\begin{center}
{\bf {\Large 
Distinguishing bounce and inflation via quantum signatures from cosmic microwave background
}}
\end{center}
\begin{center}
\vskip 0.2in
{\bf S. Mahesh Chandran} and 
{\bf S. Shankaranarayanan}\\

{\it Department of Physics, Indian Institute of Technology Bombay, Mumbai 400076, India} \\
{\tt Email: maheshchandran@iitb.ac.in, shanki@iitb.ac.in}\\
\end{center}

\vspace*{0.5cm}

\begin{abstract}
\end{abstract}
Cosmological inflation is a popular paradigm for understanding Cosmic Microwave Background Radiation (CMBR); however, it faces many conceptual challenges. An alternative mechanism to inflation for generating an almost scale-invariant spectrum of perturbations is a \emph{bouncing cosmology} with an initial matter-dominated contraction phase, during which the modes corresponding to currently observed scales exited the Hubble radius. Bouncing cosmology avoids the initial singularity but has fine-tuning problems. Taking an \emph{agnostic view} of the two early-universe paradigms, we propose a quantum measure --- Dynamical Fidelity Susceptibility (DFS) of CMBR --- that distinguishes the two scenarios. Taking two simple models with the same power-spectrum, we explicitly show that DFS behaves differently for the two scenarios. We discuss the possibility of using DFS as a distinguisher in the upcoming space missions. 
\vspace*{2.0cm}

\begin{center}
{\bf Essay received honorable mention in Gravity Research Foundation essay competition 2024.}
\end{center}
\end{titlepage}

\baselineskip=18pt


The discovery of Cosmic Microwave Background radiation (CMBR) stands as a significant milestone in our understanding of the Universe. Initially proposed as a relic of the hot big bang model by Gamow, Alpher, and Herman in the late 1940s~\cite {1946-Gamow-PR,1948-Alpher.Herman-Nature,1949-Alpher.Herman-PR}, its theoretical existence became a reality when technological progress enabled Penzias and Wilson to observe the persistent microwave background noise in all directions~\cite{1965-Penzias.Wilson-ApJ}. Subsequent measurements of the CMBR at various wavelengths have improved the accuracy of the radiation temperature to $2.73~{\rm K}$~\cite{Mather:1990tfx}. The observation of CMBR became crucial in differentiating between the steady-state and Big Bang theories of cosmology. 

To explain the large-scale structures observed in the current Universe, the Big Bang theory necessitates some level of \emph{clumpiness} in the early-universe. 
In the early 1970s, Harrison and Zeldovich argued from various perspectives that the primordial matter power spectrum should be scale-invariant~\cite{1970-Harrison-PRD,1972-Zeldovich-MNRAS}. A significant development occurred in 1980 when Fabbri et al. highlighted that CMBR anisotropies at angular scales larger than a few degrees would encompass the horizon at the last scattering surface and provide a new upper bound on the fluctuations 
~\cite{1980-Fabbri.etal-PRD,1980-Fabbri.etal-PRL}. There was, however, no  mechanism for generating initial perturbations across astronomical scales at that time, making scale-invariance a suitable choice.

In the 1980s and 1990s, two mechanisms were often considered to generate nearly scale-invariant power-spectrum --- \emph{cosmological inflation}, where quantum fluctuations expand to astronomical scales due to accelerated expansion~\cite{1980-Starobinsky-PLB,1981-Guth-PRD,1981-Sato-MNRAS,1982-Linde-PLB,1982-Linde_2-PLB,1982-Albrecht.Steinhardt-PRL}, and \emph{spontaneous symmetry breaking}, which involves relics of topological defects (such as cosmic strings) from a higher-energy space vacuum state~\cite{1987-Brandenberger-IJMPA,1991-Brandenberger-PhysScripta}. Each of these mechanisms carries implications for the resulting density fluctuation spectra. With advancements in cryogenic technology in space~\cite{1994-Benoit.Pujol-Cryo,Triqueneaux2006288}, 
measurements of CMBR temperature fluctuations by missions like WMAP~\cite{2003-WMAP} and PLANCK~\cite{Planck-2018} have \emph{ruled out} cosmic strings as the primary source of perturbations~\cite{2005-Jeong.Smoot-ApJ}. 

While \emph{single-field inflationary models} are phenomenologically successful and consistent with WMAP and PLANCK~\cite{2018-Planck-Inflation, Ellis:2023wic}, inflation faces important conceptual challenges~\cite{Steinhardt:2004rf,steinhardt2011inflation, Ijjas:2013vea, Brandenberger:2016uzh,Garfinkle:2023vzf}. Many toy models of inflation have been proposed, but embedding inflation into a realistic particle-physics model has been a challenge~\cite{1999-Lyth.Riotto-PRep}. Additionally, inflation does not remove the initial Big Bang singularity, prompting the question: Is there a new model of the early-universe that can address some of the issues with inflation while maintaining its observational successes? \emph{Matter bounce paradigm} avoids the initial big bang singularity at a classical level by violating Strong Energy Condition (SEC) in General Relativity (GR) or modify gravity by including other matter fields~\cite{Novello:2008ra,Cattoen:2005dx,Battefeld:2014uga,Nojiri:2017ncd,Ijjas:2018qbo}. Thus, in this scenario, the \emph{matter bounce} replaces the big bang, where the Universe contracts, since the infinite past, towards the bounce, after which it transitions to an expanding phase.

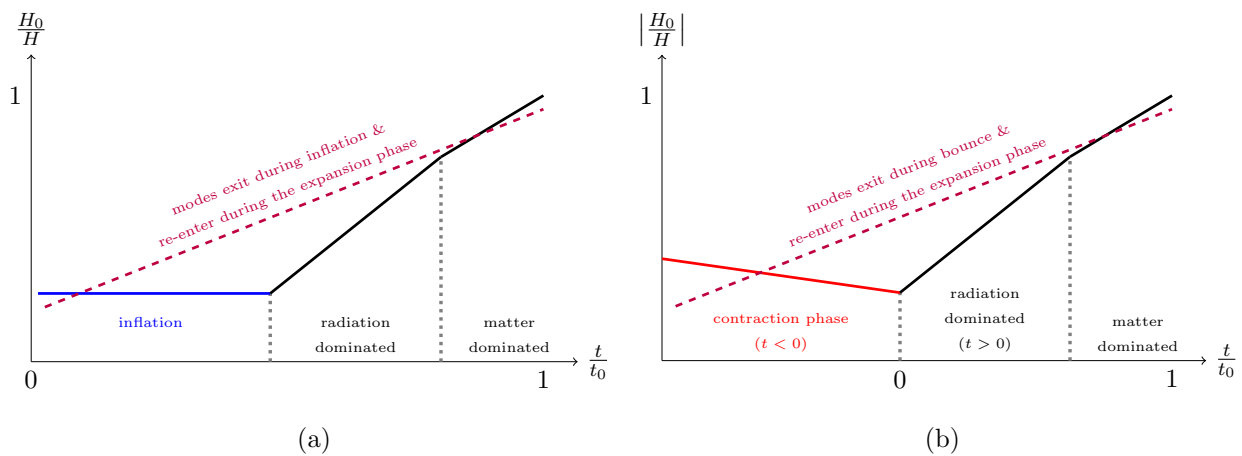
\begin{figure*}[!ht]
		\begin{center}
			\subfloat[\label{tl1}][]{\resizebox{0.5\textwidth}{!}{\begin{tikzpicture}
  \draw[->] (0, 0)node[below] {$0$} -- (8, 0)node[right] {$\frac{t}{t_0}$}  ;
  \draw[->] (0, 0) --(0, 4.5) node[above] {$\frac{H_0}{H}$};
  \draw[line width=0.5mm, dotted, gray](3.5, 0)-- (3.5, 1)  ;
   \draw [line width=0.4mm,blue] (0.1,1)--(3.5,1) ;
    \draw[line width=0.4mm,black] (3.5,1)--(6,3);
   \node[above,black,font=\tiny,align=center] at (4.75,0) {radiation\\dominated} ;
    \draw[line width=0.4mm,black] (6,3)--(7.5,3.9);
    \draw[line width=0.5mm, dotted, gray](6, 0)-- (6, 3) ;
    \node[above,blue,font=\tiny,align=center] at (1.75,0) {inflation\\};
    \node[above,black,font=\tiny,align=center] at (7,0) {matter\\dominated};
    \node[below] at (7.5,0) {$1$};
    \node[left] at (0,3.9) {$1$};
  \draw[line width=0.4mm, dashed, purple] (0.2,0.8)--node[above,sloped,purple,font=\tiny,align=center] {modes exit during inflation \&\\re-enter during the expansion phase}(7.5,3.7);
\end{tikzpicture}}
			}
			\subfloat[\label{tl2}][]{\resizebox{0.5\textwidth}{!}{\begin{tikzpicture}
  \draw[->] (0, 0) -- (8, 0)node[right] {$\frac{t}{t_0}$} ;
  \draw[->] (0, 0) --(0, 4.5) node[above] {$\abs{\frac{H_0}{H}}$};
  \draw[line width=0.5mm, dotted, gray](3.5, 0)node[below,black]{$0$}-- (3.5, 1)  ;
   
   \draw [line width=0.4mm,red] (0,1.5)--(3.5,1);
    \draw[line width=0.4mm,black] (3.5,1)--(6,3);
   \node[above,black,font=\tiny,align=center] at (4.75,0) {radiation\\dominated\\$(t>0)$} ;
    \draw[line width=0.4mm,black] (6,3)--(7.5,3.9);
    \draw[line width=0.5mm, dotted, gray](6, 0)-- (6, 3) ;
    \node[above,red,font=\tiny,align=center] at (1.75,0) {contraction phase\\$(t<0)$};
    \node[above,black,font=\tiny,align=center] at (7,0) {matter\\dominated};
    \node[below] at (7.5,0) {$1$};
    \node[left] at (0,3.9) {$1$};
  \draw[line width=0.4mm, dashed, purple] (0.2,0.8)--node[above,sloped,purple,font=\tiny,align=center] {modes exit during bounce \&\\re-enter during the expansion phase}(7.5,3.7);
\end{tikzpicture}}
			}
\caption{Plot of Hubble radius $H^{-1}$ as a function of cosmic time $t$. Both Hubble radius and cosmic time are rescaled such that their current value is $1$.  The left plot depicts a Universe with an initial period of accelerated expansion. The right plot depicts the Universe $(t > 0 )$ evolved from a contracting matter dominated phase $(t < 0 )$ to a bounce at $t = 0$.}
\label{fig:timeline}
		\end{center}
	\end{figure*}

In both scenarios, quantum vacuum oscillations in the matter and the gravitational fields lead to classical fluctuations in the energy density. These fluctuations serve as the seeds for temperature anisotropies and polarization in the CMBR, as well as for the formation of Large Scale Structures in the present Universe~\cite{Novello:2008ra,Battefeld:2014uga,Ijjas:2018qbo}. From \ref{fig:timeline}, we see that both inflation and matter bounce scenarios provide a causal mechanism for the generation of primordial perturbations that has observable imprints on the CMBR~\cite{Battefeld:2014uga,Nojiri:2017ncd,Ijjas:2018qbo}. Precisely, in the inflationary scenario, the Hubble radius, or the “apparent” horizon (blue curve), and the forward light cone (purple dotted curve), or the ``actual horizon," do not coincide. In the bouncing scenario, the Hubble radius (red curve) shrinks during the contraction phase. At the level of the power spectrum, a duality between contracting phase and inflating Universe exists~\cite{1998-Wands-PRD,2002-Tsujikawa-PLB,2002-Finelli.Brandenberger-PRD}. Specifically, it is shown that a contracting Universe dominated by pressure-less cold matter gives rise to scale-invariant power-spectrum~\cite{1998-Wands-PRD,2002-Tsujikawa-PLB,2002-Finelli.Brandenberger-PRD}. 

Taking an \emph{agnostic view} of the two early-universe paradigms, we ask the following questions: Can CMBR provide key measures to distinguish between inflation and bounce? Can quantum signatures provide a route to distinguish them? Due to the duality relation~\cite{1998-Wands-PRD}, any measure proportional to the power spectrum will lead to a null test. Hence, the new measures must surpass the current standards for extracting quantum signatures from CMBR~\cite{2003-WMAP,Planck-2018}. Given that the quantum vacuum oscillations in the matter and the gravitational fields lead to classical fluctuations in the energy density, we propose a quantum measure that can distinguish between the two scenarios. While the exact conditions that steered the early moments of our Universe may now be well beyond our reach, we show in this essay that the \textit{dynamical fidelity susceptibility} (DFS) distinguishes the two scenarios~\cite{2007YouPhys.Rev.E,gu2010fidelity,2014Strobel-Science,2016HaukeNaturePhysics,2019KattemoellePhys.Rev.A,2019-Richards-Mas}.


To go about this, we consider a massless scalar field in Friedmann-Lemaitre-Robertson-Walker space-time whose action is given below~\cite{1984Kodama.SasakiProg.Theor.Phys.Supp.,1992Mukhanov.etalPhys.Rept.}:
\begin{equation}
S=\frac{1}{2}\int d^4x \sqrt{-g}g^{\mu\nu}\partial_{\mu}\Phi\partial_{\nu}\Phi\quad;\quad ds^2=g_{\mu\nu}dx^{\mu} dx^{\nu}=dt^2-a^2(t)(dr^2+r^2d\Omega^2),
\end{equation}
where $a(t)$ is the scale factor. At the linear level, the 
equations describing both gravitational and matter perturbations can be quantized consistently via gauge-invariant variables. The scalar (density) perturbations are given by Mukhanov-Sasaki variables~\cite{1984Kodama.SasakiProg.Theor.Phys.Supp.,1992Mukhanov.etalPhys.Rept.}, while the tensor perturbations are the gravitational waves. The duality transformation between matter dominated collapse and inflating Universes works for both sets of perturbations~\cite{1998-Wands-PRD}.

Upon employing a partial-wave decomposition in terms of real spherical harmonics ($l,m$), followed by lattice-regularization of the field along the co-moving radial direction ($r=jd$), we obtain the scalar-field Hamiltonian in terms of the canonically conjugate field variables $\{\Pi_{lmj},\Phi_{lmj}\}$ as follows~\cite{2024Chandran.etalPhys.Rev.D}:
\begin{equation}
   \mathcal{H}(t)=\frac{1}{2}\sum_{lmj}\left[\Pi_{lmj}^2+\sum_k K^{(lm)}_{jk}(t)\Phi_{lmj}\Phi_{lmk}\right],
   \label{eq:Hamiltonian}
\end{equation}
which resembles a network of time-dependent oscillators, where the coupling matrix $K_{jk}^{(lm)}$ has a tridiagonal form indicative of nearest-neighbour coupling between the oscillators. Here, the lattice-spacing $d$ fixes the UV-cutoff and radial boundary $Nd$ fixes the IR-cutoff. Also, it should be noted that all quantities in the above Hamiltonian have already been rescaled to be dimensionless, for e.g., cosmic-time $t$ here replaces $t/d$ and Hubble parameter $H$ replaces $Hd$ in the derivation~\cite{2023Chandran.ShankaranarayananPhys.Rev.D,2024Chandran.etalPhys.Rev.D}. Diagonalizing this matrix allows the Hamiltonian to be decoupled in terms of a time-dependent normal mode spectrum $\{\omega_{k}^2(t)\}$. The wave-function evolution for each normal mode is highly sensitive to the sign of $\omega_k^2$, which depends on whether the mode is inside ($\omega_k^2>0$) or outside ($\omega_k^2<0$) the (dimensionless) Hubble radius ($H^{-1}$)~\cite{2023Chandran.ShankaranarayananPhys.Rev.D,2024Chandran.etalPhys.Rev.D}. 

To identify the difference in the two early-universe scenarios, at the initial time, we assume all the quantum fluctuations to be sub-Hubble ($\omega_k^2>0$) and in the ground state for both scenarios~\cite{2024Chandran.etalPhys.Rev.D,2023Chandran.ShankaranarayananPhys.Rev.D}. Since the Hamiltonian \eqref{eq:Hamiltonian} is quadratic, the wave-function $\ket{\Psi}$ retains a Gaussian form throughout the evolution~\cite{2023Chandran.ShankaranarayananPhys.Rev.D}. While the overall system remains in a pure state, i.e., $\Tr\rho^2=1$, its constituent subsystems, corresponding to field degrees of freedom confined to spatial subregions, are in a mixed state. The subsystem corresponding to a subregion $A$ is therefore described by a reduced density matrix (RDM) $\rho_A$ obtained by tracing out the complementary degrees of freedom (say, $B=A'$) from the overall density matrix as $\rho_A=\Tr_B \ket{\Psi}\bra{\Psi}$. The spatial subregions of the field are therefore \textit{entangled}, the extent of which can be quantified with the help of von Neumann entropy $S_A=-\Tr\rho_A\ln\rho_A$. Here, the total entanglement entropy of the subsystem is the sum over all angular mode contributions $S=\sum_l(2l+1)S_l$, which converges for large $l$~\cite{1993-Srednicki-Phys.Rev.Lett.}.

To see the effects of an expanding Universe on the \emph{quantumness} of CMBR, we may focus on the off-diagonal elements of the corresponding RDM. When these elements fall to zero, the RDM resembles a classical statistical ensemble, i.e., the subsystem \emph{decoheres} upon interacting with environmental degrees of freedom. However, decoherence is, in general, basis-dependent. A stronger condition may, therefore, be imposed by requiring the RDM to be \textit{maximally mixed}, wherein it is proportional to the Identity matrix ($\rho_A\propto \mathscr{I}$), resulting in decoherence across \textit{all} bases. For infinite-dimensional Hilbert spaces such as the ones we consider here, maximally mixed states correspond to vanishing purity ($\Tr\rho_A^2\to0$), or equivalently, a diverging entanglement entropy ($S_A\to\infty$). Recently~\cite{2024Chandran.etalPhys.Rev.D,2023Boutivas.etal}, it was shown that inflationary models resulted in the rapid growth of entanglement entropy in spatial subregions, indicating \textit{self-decoherence} of fluctuations due to the expansion~\cite{2021Martin.VenninJournalofCosmologyandAstroparticlePhysics}. 

While the entanglement entropy can shed some light on the non-trivial effects due to a time-dependent background, it has its limitations. For instance, it is known that the Hamiltonian for linearized fluctuations in conformal time is invariant under a duality transformation~\cite{1998-Wands-PRD}, resulting in two models with identical power spectra. This implies that, regardless of the time-coordinate used, the ensuing dynamics of both models starting from the same vacuum state cannot be distinguished by entanglement entropy as it is a symplectic invariant (preserved under canonical transformations)~\cite{2024Chandran.etalPhys.Rev.D}. 
This limits its applicability to distinguish between early-universe models compared to other quantum measures, as discussed below.

\begin{figure}[!hbt]
	\centering
	\includegraphics[scale=0.45]{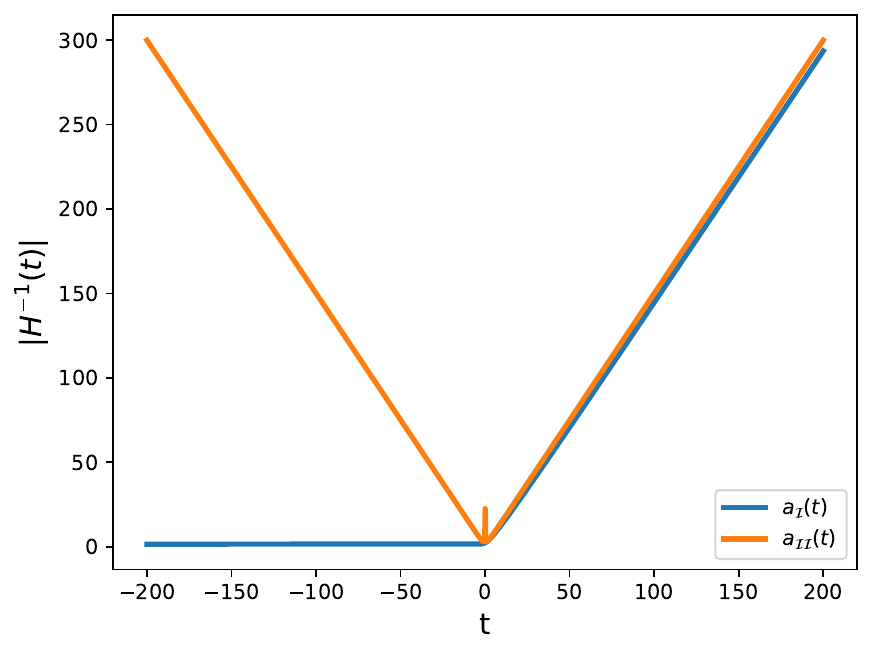}
	\caption{Hubble radius $\abs{H^{-1}}$ for the two models with scale-factors $a_I(t), a_{II}(t)$ in Eq.~\eqref{eq:twomodels}.
 We have set $H_0=1$ and $t_e=0$.}
	\label{fig:Hrad}
\end{figure}

\begin{figure*}[!ht]
		\begin{center}
			\subfloat[\label{ent1}][]{%
				\includegraphics[width=0.44\textwidth]{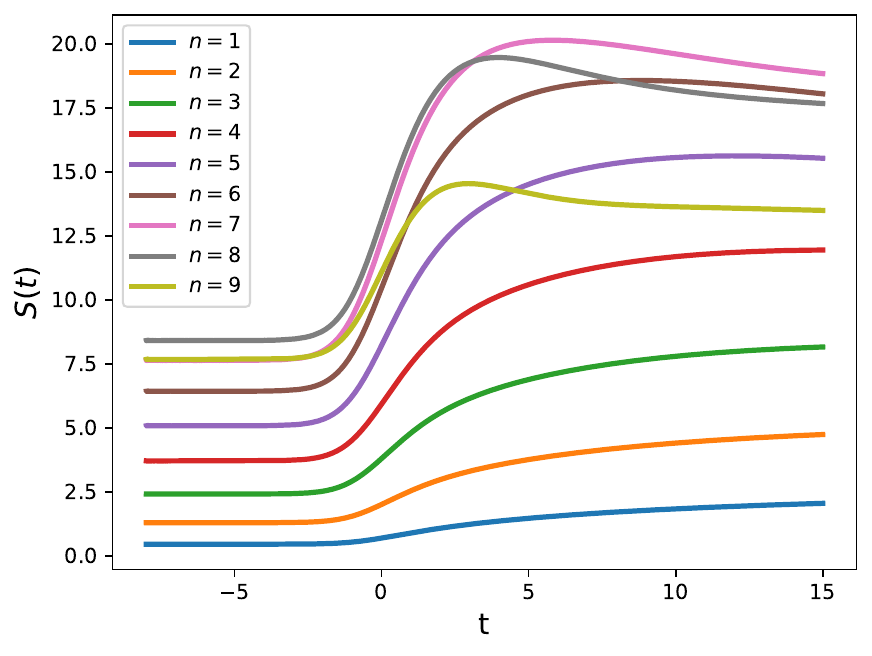}
			}
			\subfloat[\label{ent2}][]{%
				\includegraphics[width=0.43\textwidth]{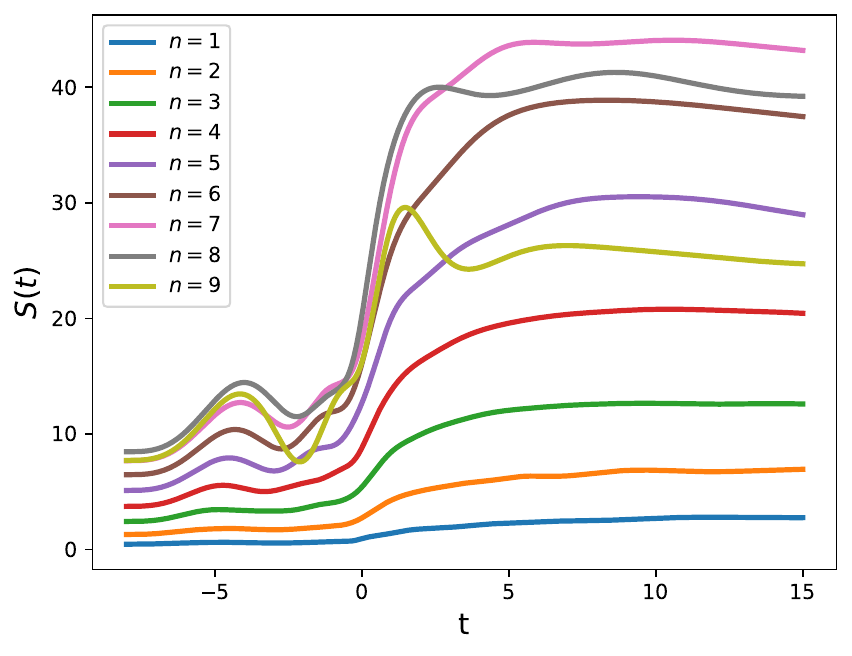}
			}
			
			\caption{Entanglement entropy of massless scalar field for (a)  $a_I(t)$: inflation followed by matter-dominated expansion, and (b) $a_{II}(t)$: matter-bounce. Here, $N=10$, and we count angular mode contributions up to $l=10$. 
   While the dynamics are expected to coincide at earlier time-points $t \to-\infty$, starting closer to the singularity ($a_I \to 0$) also requires tremendous computing power.}
			\label{fig:ent}
		\end{center}
	\end{figure*}

To keep things transparent, we consider two simple models for the two early-universe scenarios --- i) inflation followed by matter-dominated expansion $a_I(t)$, and ii) matter-bounce $a_{II}(t)$ (cf. \ref{fig:Hrad}):
\begin{equation}
\label{eq:twomodels}
\!\! a_I(t)=\left[\mathcal{W}_0\left(e^{H_0(t-t_e)}\right)\right]^{2/3}\sim\begin{cases}
    e^{\frac{2}{3}H_0(t-t_e)}& t\ll t_e\\
    \left(H_0(t-t_e)\right)^{2/3}& t\gg t_e
    \end{cases}~;~a_{II}(t)=\left(1+H_0^2(t-t_e)^2\right)^{1/3},
\end{equation}
where $\mathcal{W}_0(x)$ is the principal branch of the Lambert function, and $t_e$ is the transition time for de Sitter expansion during inflation to matter-dominated expansion in $a_I(t)$ and matter-dominated collapse to matter-dominated expansion in $a_{II}(t)$. The reasons for these choices are two-fold --- i) it enables semi-analytical evaluation of the quantum measures we are interested in, based on techniques recently outlined by the current authors~\cite{2023Chandran.ShankaranarayananPhys.Rev.D,2024Chandran.etalPhys.Rev.D}, and ii) the initial phases for both models, i.e., de Sitter expansion in $a_I$ and matter-dominated collapse in $a_{II}$ correspond to identical power-spectra~\cite{1998-Wands-PRD}, which will help us better identify the quantum measure that can clearly isolate the differences in the two early-universe scenarios. While inflation begins after $t>0$ in cosmic time, we have shifted its origin such that the time at which both models $a_I(t)$ and $a_{II}(t)$ transition to radiation-dominated expansion coincide $(t_e)$. In both models, the modes exit the Hubble radius during inflation/bounce and later re-enter during the matter-dominated expansion phase (cf. \ref{fig:timeline}). Since we can only observe up to the largest wavelength mode that re-enters, it may be difficult to tell apart whether inflation or bounce preceded the expansion. From \ref{fig:ent}, we also see that the qualitative features of entanglement dynamics are largely similar for both models. In fact, since inflation and matter-dominated collapse are duals~\cite{1998-Wands-PRD}, the entanglement dynamics for $t<0$ in both cases are expected to coincide upon evolving them from the same vacuum state~\cite{2024Chandran.etalPhys.Rev.D}. Lastly, since the above scale factors are continuous, spurious effects that may otherwise arise in quantum measures have also been avoided.

In order to distinguish the two classes of early-universe models, it is ideal to rely on quantum measures that are highly sensitive to the ensuing dynamics. One fundamental quantity that is both conceptually rich and experimentally accessible is the \textit{dynamical fidelity susceptibility} (DFS)~\cite{2007YouPhys.Rev.E,gu2010fidelity,
2014Strobel-Science,2016HaukeNaturePhysics,2019KattemoellePhys.Rev.A,2019-Richards-Mas}. Mathematically, they are evaluated from the dynamical fidelity~\cite{1984PeresPhys.Rev.A,2016WimbergerPTRSA} defined for pure (overall system) and mixed states (subsystems) as follows:
\begin{equation}
    \mathscr{F}_{\text{pure}}=\bra{\Psi(t)}\ket{\Psi(t+\delta t)}\quad;\quad \mathscr{F}_{\text{mixed}}=\Tr\sqrt{\rho^{1/2}(t)\rho(t+\delta t)\rho^{1/2}(t)} \, .
\end{equation}
 Expanding the dynamical fidelity about small values of $\delta t$, we get:
\begin{equation}
    \mathscr{F}\sim 1-\chi_F(t)\delta t^2\quad;\quad \chi_F=\frac{1}{2}\frac{d^2\mathscr{F}}{d (\delta t)^2}\bigg\rvert_{\delta t=0},
\end{equation}
where the first derivative vanishes since fidelity is maximum (unity) at $\delta t=0$, and the second derivative ($\chi_F$) determines the DFS. 
In this essay, we use cosmic time $t$ as the control parameter, whereas \textit{fidelity} and \textit{fidelity susceptibility} are more generally defined based on some control parameter depending on the physical system at hand.
We therefore add the prefix ``dynamical" everywhere owing to this choice~\cite{2019KattemoellePhys.Rev.A,2019-Richards-Mas}. The DFS therefore captures important features exclusive to system dynamics, some of which are given below:
\begin{itemize}
    \item For pure states, it is related to the dynamical quantum Fisher information $F_Q(t)$ and satisfies the following quantum Cramer-Rao bound:
    \begin{equation}\label{eq:qfi}
        \left(\Delta t\right)^2\geq \frac{1}{F_Q(t)}\quad;\quad F_Q(t)=\frac{\chi_F(t)}{2},
    \end{equation}
    which in this case fixes the maximum precision for estimating \textit{time scales} from the state evolution~\cite{1994BraunsteinPhys.Rev.Lett.,2019-Richards-Mas}. It also coincides with the \textit{quantum speed limit} (QSL), which fixes the minimum time it takes for the state $\rho(t)$ to unitarily evolve to a nearby, distinguishable state $\rho(t+\delta t)$ in response to the conditions that drive the evolution~\cite{2013TaddeiPhys.Rev.Lett.,2016PiresPhys.Rev.X,2018GessnerPhys.Rev.A}.
    \item It captures potential signatures of \textit{dynamical quantum phase transitions} (DQPT) in the thermodynamic limit $N\to\infty$ of many-body systems~\cite{2006Zanardi.PaunkoviifmmodeacutecelsecfiPhys.Rev.E,2008Zhou.BarjaktarevicJournalofPhysicsAMathematicalandTheoretical,2010VieiraJournalofPhysicsConferenceSeries,2019-Richards-Mas}. The idea rests on the fact that at transition points, a small change in the control parameter (here, cosmic time) results in greatly enhanced distinguishability of corresponding states, which would therefore be reflected in the DFS~\cite{2007ZanardiPhys.Rev.Lett.}.
    \item The QFI density $f_Q=F_Q/N$ is a useful measure for multi-partite entanglement in $N$-body systems, i.e., the system is $(m+1)$-\textit{partite entangled} if $f_Q> m$~\cite{2012HyllusPhys.Rev.A,2012TothPhys.Rev.A,2016HaukeNaturePhysics}.
\end{itemize}

\begin{figure*}[!ht]
		\begin{center}
			\subfloat[\label{xf1}][]{%
				\includegraphics[width=0.425\textwidth]{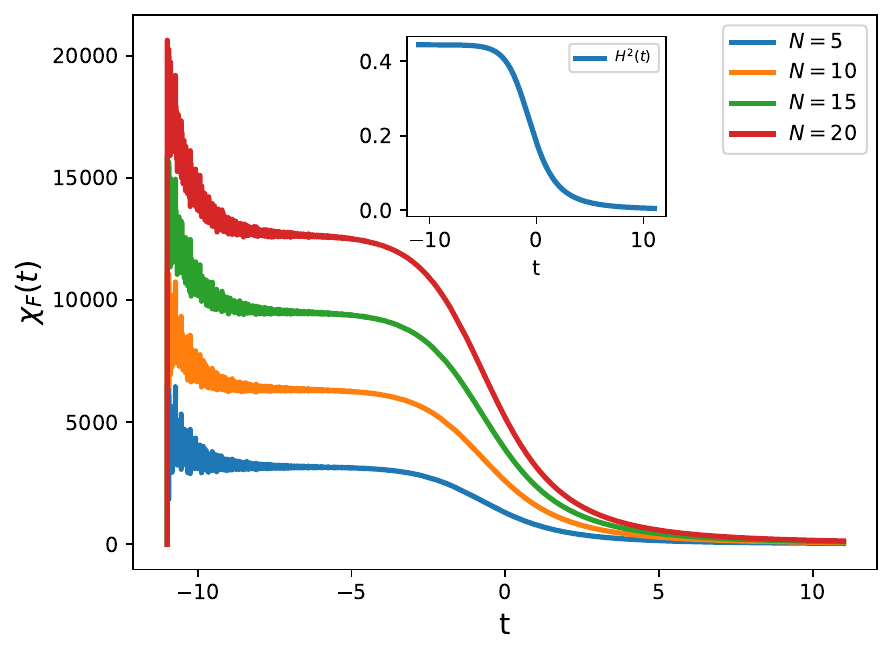}
			}
			\subfloat[\label{xf2}][]{%
				\includegraphics[width=0.425\textwidth]{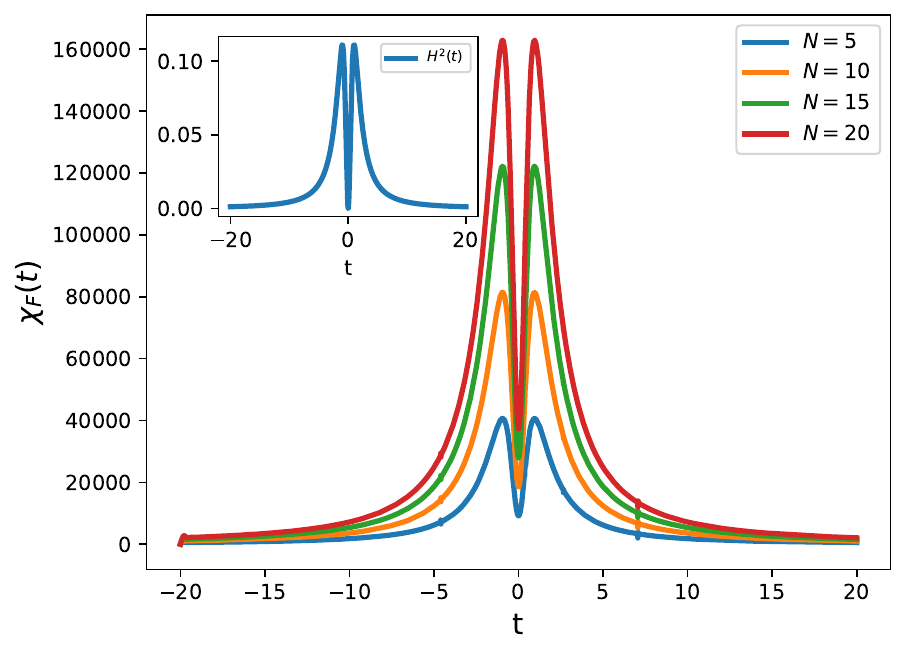}
			}
			
			\caption{Dynamical fidelity susceptibility for the overall state of the fluctuation when the background undergoes (a) inflation followed by matter-dominated expansion and (b) a matter-bounce. Here, we count angular mode contributions up to $l=150$ for (a) and $l=1000$ for (b).}
			\label{fig:xfN}
		\end{center}
	\end{figure*}

 \begin{figure*}[!ht]
		\begin{center}
			\subfloat[\label{xf1}][]{%
				\includegraphics[width=0.425\textwidth]{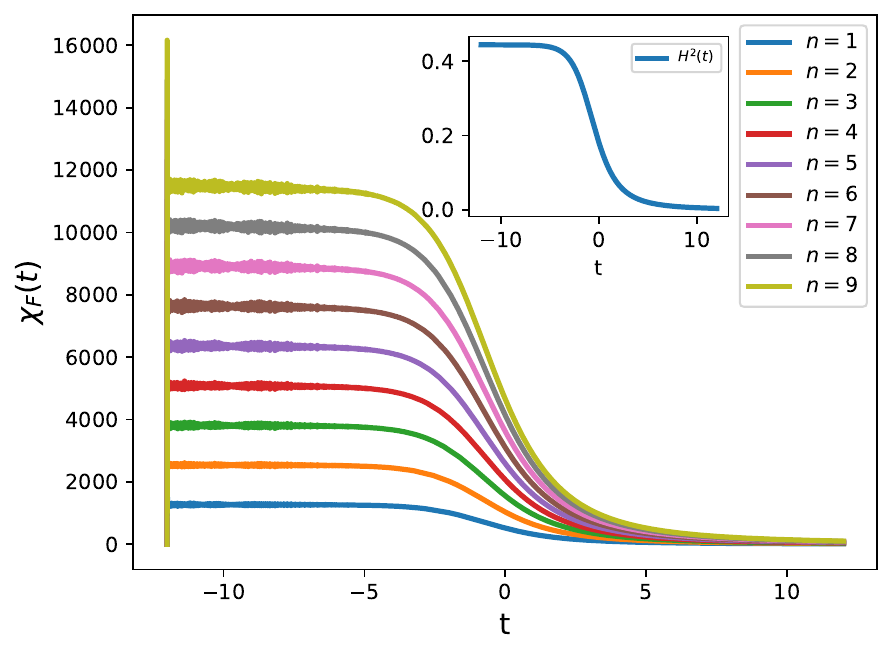}
			}
			\subfloat[\label{xf2}][]{%
				\includegraphics[width=0.425\textwidth]{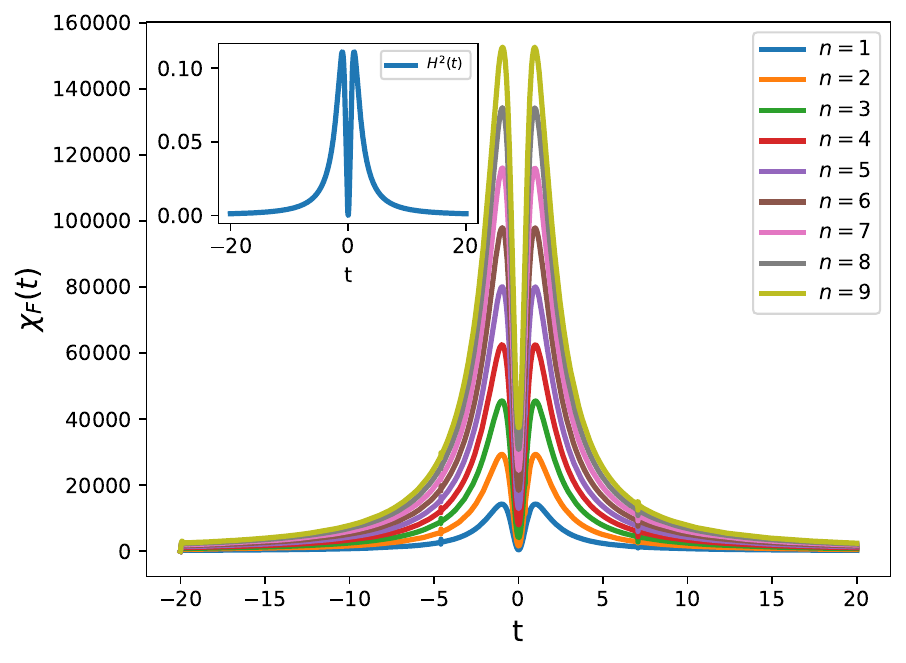}
			}
			
			\caption{Spatial subregion dynamical fidelity susceptibility of fluctuations when the background undergoes (a) inflation followed by matter-dominated expansion and (b) a matter-bounce. Here, $N=10$, and we count angular mode contributions up to $l=150$ for (a) and $l=1000$ for (b).}
			\label{fig:xfn}
		\end{center}
	\end{figure*}

To isolate model-specific features arising from cosmological expansion, we simulate the DFS evolution in \ref{fig:xfN} for the overall state $\ket{\Psi(t)}$ corresponding to various system sizes ($N$). One can see that the distinguishing feature is the existence of \textit{two distinct peaks} in the DFS evolution for the bounce model that are absent in the inflation model at the transition point $t=0$. The overall behavior is also found to match that of the Hubble parameter as $\chi_F\propto H^2$. It can also be seen that the peaks observed in the case of bounce increase with system size $N$, which can be extrapolated to a possible divergent behavior in the thermodynamic limit $N\to\infty$. The DFS evolution, therefore, hints at DQPT in the bouncing model, contrary to the inflationary counterpart. In \ref{fig:xfn}, we simulated the DFS for spatial subregions and observed similar peaks in the bouncing model that were absent in the inflationary model. The peaks also became more pronounced with subsystem size $n$, implying that spatial subregions also could be subject to DQPT in the thermodynamic limit $n\to\infty$ (such that $n<N$). Although the inflation plots indicate a \textit{separate} peak 
at the start of the evolution, this may be an artefact of the numerical limitations on the time step $(\delta t)$ required to sufficiently resolve the wave-function evolution close to the singularity $(a_I\to 0)$. While $a_I(t)$ and $a_{II}(t)$ are simplified models, our results imply that DFS is, in general, highly sensitive to the background dynamics and can isolate features that help discriminate between early-universe models.

\begin{figure*}[!ht]
		\begin{center}
			\subfloat[\label{xf1}][]{%
				\includegraphics[width=0.425\textwidth]{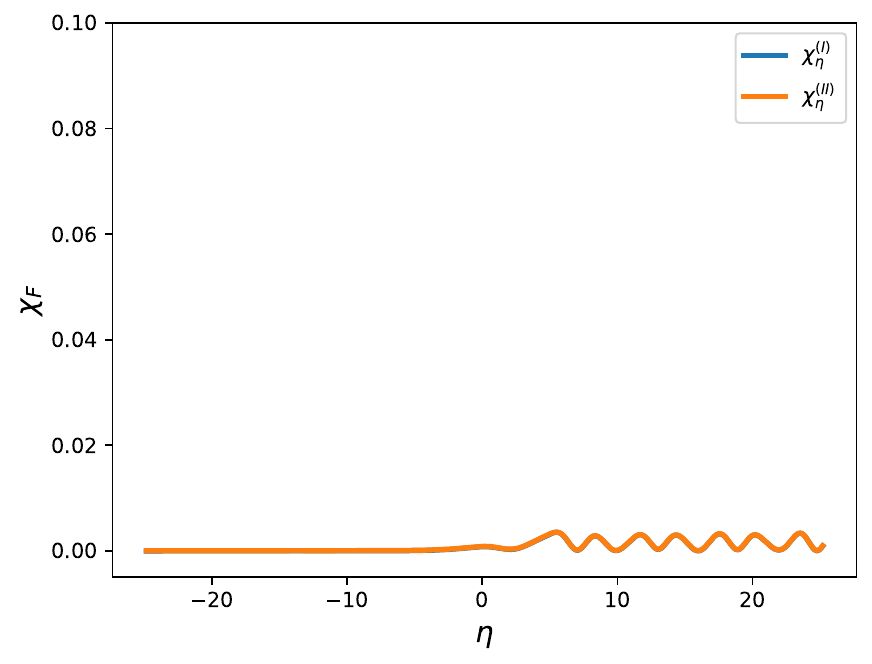}}
			\subfloat[\label{xf2}][]{%
				\includegraphics[width=0.41\textwidth]{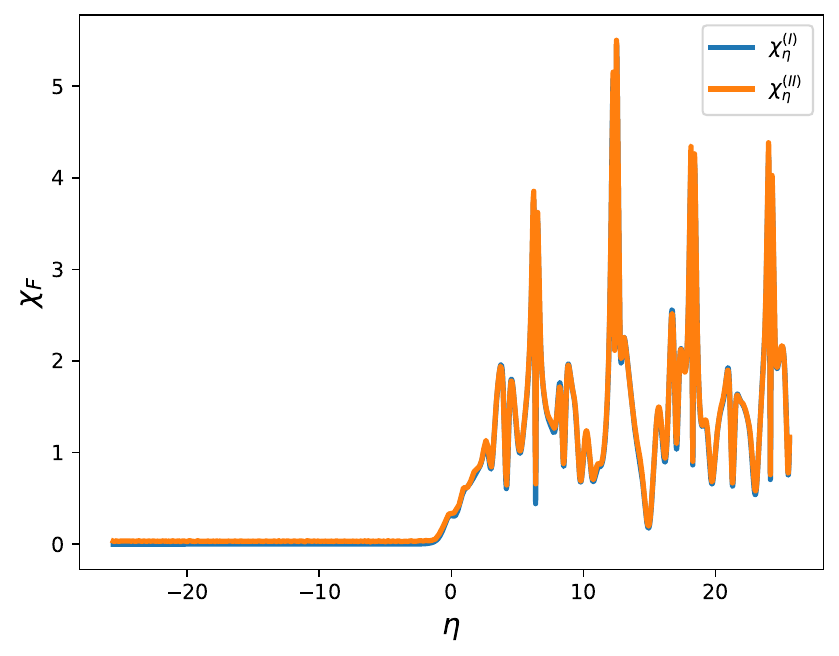}
			}
			
			\caption{Conformal time DFS ($\chi_{\eta}$) when the background undergoes (a) inflation followed by matter-dominated expansion, and (b) a matter-bounce. We see that the DFS $\chi_{\eta}^{(I)}$ obtained via conformal-time Hamiltonian, matches with the DFS $\chi_{\eta}^{(II)}$ obtained via cosmic-time Hamiltonian using the relation \eqref{eq:dfsconformal}. Here, $N=5$, and we count angular mode contributions up to $l=20$.}
			\label{fig:xfconformal}
		\end{center}
	\end{figure*}

While DFS here is defined with respect to cosmic time, the question arises as to whether such distinguishable features in its dynamics are robust to the choice of time coordinate. Here, the evolution of the Gaussian state can be fully characterized by the scaling parameters $\{b_{jlm}(t)\}$ that satisfy the Ermakov-Pinney equation corresponding to each field mode~\cite{2024Chandran.etalPhys.Rev.D}. The DFS for the overall system can then be decomposed as:
\begin{equation}
    \chi_{t}(t)=\sum_{jlm}\frac{\dot{b}_{jlm}^2(t)}{4b_{jlm}^2(t)}\equiv\sum_{jlm}x_{jlm}^2(t)\quad;\quad \chi_{\eta}(\eta)=\sum_{jlm}\frac{B_{jlm}'^2(\eta)}{4B_{jlm}^2(\eta)},
\end{equation}
where $\{b_{jlm}(t)\}$ and $\{B_{jlm}(\eta)\}$ are scaling parameters in cosmic time and conformal time, respectively. When the initial states corresponding to both choices match exactly (this corresponds to taking initial time points $t_0\to-\infty$ and $\eta_0\to -\infty$ for defining the vacuum), the scaling parameters and the DFS evolutions are related as follows~\cite{2024Chandran.etalPhys.Rev.D}:
\begin{equation}\label{eq:dfsconformal}
\frac{\dot{b}_{jlm}^2}{b_{jlm}^2}=\frac{1}{a^2(t)}\left[\frac{B'_{jlm}}{B_{jlm}}+\frac{\dot{a}(t)}{2}\right]^2\quad \Rightarrow\quad{X_{\eta}[\eta(t)]} = a^2(t) \sum_{jlm}\left[ x_{jlm}(t) - \frac{H(t)}{2} \right]^2.
\end{equation}
From \ref{fig:xfconformal}, for inflation followed by matter-dominated expansion, we see that the DFS is negligible ($\chi_{\eta}\sim 10^{-3}$) throughout, implying that nearby states strongly overlap throughout the evolution. On the other hand, for matter-bounce, we observe that multiple sharp peaks emerge from the bounce point, though not as pronounced as the peaks in cosmic time DFS but significantly more in number. These differences manifest because, as mentioned above (after Eq.~\eqref{eq:qfi}), a particular choice of the time-coordinate would correspond to a particular value of ``minimum time" it will take for a state to evolve to a nearby, distinguishable state in response to the driving conditions. Nevertheless, generalizing \eqref{eq:dfsconformal} using the approach laid out in \cite{2024Chandran.etalPhys.Rev.D}, we may shift between any such choices and identify corresponding DFS features that distinguish between inflation and bounce. For instance, for the choices of conformal time and cosmic time coordinates, inflation followed by matter-dominated expansion does not produce sharp peaks in the DFS, as opposed to matter-bounce, where such peaks manifest.

In the laboratory setting, the protocol for obtaining DFS involves preparing multiple copies of the same state $\rho(t)$ and making measurements at different time-slices $\{t+\delta t,t+2\delta t,..\}$ of its evolution~\cite{2014Strobel-Science}. In the case of Gaussian states, it is sufficient to measure the two-point correlators of conjugate variables (the covariance matrix), which has proved to be a successful feat in cold-atom quantum field simulators via tomographic methods~\cite{2020GluzaCommunicationsPhysics}. In the cosmological scenario, however, reconstructing the entire covariance matrix from the CMBR poses a considerable challenge. Nevertheless, advances have been made in utilizing the power spectrum to construct reduced covariance matrices corresponding to \textit{disjoint spatial sub-regions} of the field~\cite{2021Martin.VenninPhys.Rev.D,2021Martin.VenninJournalofCosmologyandAstroparticlePhysics,2023AgulloPhys.Rev.D}. Therefore, this approach can allow us to study the subsystem DFS from CMBR observations and better discriminate between early-universe models, owing to its high sensitivity to background dynamics. 

To conclude, advances in cosmology and particle physics are intimately linked to technological progress through a mutually beneficial cause-and-effect relationship. From the serendipitous discovery by Penzias and Wilson to the recent PLANCK mission, CMBR has provided deep insights into the primordial conditions after the Big Bang, seeds of the structure formation that lead to galaxies, allowing us to probe different epochs of the Universe with high-precision, and validating the current acceleration of the Universe~\cite{Smoot:2007zz,Staggs:2018gvf,Durrer:2020fza,Komatsu:2022nvu}. This progress has been made possible by advancements in detector and open-cycle cryogenics technology~\cite{2022-Zu.etal-Cryo}. 

This essay proposes a new quantum measure --- \emph{Dynamical Fidelity Susceptibility} --- of CMBR to distinguish between the two early-universe paradigms for structure formation currently under {popular} consideration. However, quantum measurements are highly susceptible to environmental changes. Minor fluctuations in temperature or stray electrical or magnetic fields can disrupt quantum measurements, leading to information degradation~\cite{Preskill:2018jim}. Currently, the threshold for error rate per gate for two-qubit gates stands above $0.1\%$~\cite{2019-Google-Nature,Kandala:2018kwe}. This is expected to be better in the next decade~\cite{Redchenko:2023zsw}. 
Coupled with advancements in adiabatic demagnetization refrigeration technology for space missions~\cite{2022-Zu.etal-Cryo}, there is optimism that the \emph{dynamical fidelity susceptibility} will become experimentally accessible in the next decade, bringing us closer to understanding the origins of the Universe.

%


\noindent {\bf Acknowledgments} The authors thank Himadri Dhar, Neeraj Rajak and Sai Vinjanampathy for discussions on quantum measures. The authors thank Joseph P. Johnson, Ashu Kushwaha, Snehit Pangal and Debottam Nandi for comments on the earlier draft. SMC is supported by Prime Minister's Research Fellowship offered by the Ministry of Education, Govt. of India. The work is supported by the SERB Core Research grant (SERB/CRG/2022/002348) and SPARC MoE grant SPARC/2019-2020/P2926/SL.

\end{document}